# Link between the laws of geometrical optics and the radiative transfer equation in media with a spatially varying refractive index.

Jean-Michel Tualle

*Lab. de Phys. des Lasers (CNRS UMR 7538), Univ. Paris 13*

*99 av. J.-B. Clément, 93430 Villetaneuse, France*

**ABSTRACT:**

We proposed in a previous paper [Opt. Commun. **228**, 33 (2003)] a modified radiative transfer equation to describe radiative transfer in a medium with a spatially varying refractive index. The present paper is devoted to the demonstration that this equation perfectly works in the non-absorbing / non-scattering limit, what was contested by L. Martí-López and coworkers [Opt. Commun. **266**, 44 (2006)]. The assertion that this equation would imply a zero divergence of the rays is also commented.

Corresponding author: Tel.: +(33)-1-49-40-40-92; fax.: +(33)-1-49-40-32-00.

E-mail address: tualle@galilee.univ-paris13.fr

Radiative transfer in a medium with a spatially varying refractive index was recently at the origin of a lot of activity [1-14]. We proposed in a previous paper [3] the following modified radiative transfer equation (RTEVI) to describe such a problem, that is:

$$\frac{n(\vec{r})}{c_0}\frac{\partial}{\partial t}L(\vec{r},\vec{\Omega},t) + \vec{\Omega}\cdot\vec{\nabla}L(\vec{r},\vec{\Omega},t) - 2[\vec{\Omega}\cdot\vec{\nabla}\ln n(\vec{r})]L(\vec{r},\vec{\Omega},t)$$
$$+ \vec{\nabla}\ln n(\vec{r})\cdot\vec{\nabla}_\Omega L(\vec{r},\vec{\Omega},t) + [\mu_a(\vec{r})+\mu_s(\vec{r})]L(\vec{r},\vec{\Omega},t) \quad (1)$$
$$= \mu_s(\vec{r})\int_{4\pi}\theta(\vec{r},\vec{\Omega},\vec{\Omega}')L(\vec{r},\vec{\Omega}',t)d\Omega' + \varepsilon(\vec{r},\vec{\Omega},t)$$

where $L(\vec{r},\vec{\Omega},t)$ is the radiance, $n(\vec{r})$ the refractive index, $c_0$ the speed of light in vacuum, $\mu_a$ and $\mu_s$ the absorption and scattering coefficients, $\theta(\vec{r},\vec{\Omega},\vec{\Omega}')$ the normalized scattering function, $\varepsilon(\vec{r},\vec{\Omega},t)$ the source term, and where $\vec{\nabla}$ denotes the gradient operator with respect to coordinate $\vec{r}$, while $\vec{\nabla}_\Omega$ is the transverse gradient operator with respect to coordinate $\vec{\Omega}$ (we will come back to the nature of this last operator in the appendix).

This expression was confirmed by M. Shendeleva [4], who notes that it was in fact first derived by G.C. Pomraning [15]. Furthermore, G. Bal proved [7] that it corresponds to the high frequency limit of Maxwell equations in heterogeneous media [16]. One can add that, as it was underlined in [3], this expression satisfies the energy conservation equation: if we introduce the quantities

$$\varphi(\vec{r},t) = \int L(\vec{r},\vec{\Omega},t)d\Omega \text{ and } \vec{j}(\vec{r},t) = \int \vec{\Omega} L(\vec{r},\vec{\Omega},t)d\Omega \quad (2)$$

we indeed have the conservation equation:

$$\frac{\partial\varphi}{\partial t} + \vec{\nabla}\cdot\vec{j} + \mu_a\varphi = E(\vec{r},t) \quad (3)$$

where $E$ is a source term. We can prove this fact in the same way as for the classical RTE equation (which is well-known to satisfy this equation [17]), by integrating (1) over $\vec{\Omega}$. The demonstration is exactly the same (please note that $\vec{\Omega}$ does not depend on $\vec{r}$), except for the terms in $\vec{\nabla} \ln r$ that take into account the spatial variations of the refractive index. The first one can be readily integrated, and leads to:

$$-2\vec{\nabla} \ln r \cdot \int \vec{\Omega} L(\vec{r},\vec{\Omega},t) d\Omega = -2\vec{\nabla} \ln r \cdot \vec{j} \tag{4}$$

The second term is more subtle. T. Khan and H. Jiang proved [2] by doing the complete calculation that

$$\int \vec{\nabla}_\Omega L(\vec{r},\vec{\Omega},t) d\Omega = 2\vec{j}, \tag{5}$$

and we propose a more general demonstration of this fact in the appendix. One can therefore see that these two contributions exactly cancel, when the other terms give the conservation equation (3). There are therefore a lot of elements that attest to the validity of this modified RTE.

L. Martí-López, J. Bouza-Domínguez, R.A. Martínez-Celorio, J.C. Hebden however contest its validity [14], and argue that it would not be able to explain some well-known results in the non-scattering / non-absorbing limit. They considered two time-independent problems in this limit:

- The radiation from a point-like source in a medium with a constant refractive index. One has to find an inverse square law for the light intensity.
- The evolution of the intensity along a classical ray in a medium with a spatially varying index, where one should have [18,14]

$$\frac{d}{ds}\left[\frac{I(s)}{n(s)}\right] = -\frac{\Delta \Lambda(s)}{n^2(s)} \tag{6}$$

where $s$ denotes the arc length along the classical ray, and where $\Lambda(s) \equiv \Lambda(\vec{r}(s))$ is the eikonal.

Let us now examine how equation (1) can completely solve these problems in a satisfying manner, and let us start with the first one. In a time-independent problem, with a point-like time-independent source term $E = P_0\delta(\vec{r})$, and with $\mu_s = \mu_a = 0$, the conservation equation reads:

$$\vec{\nabla} \cdot \vec{j} = P_0 \delta(\vec{r}) \tag{7}$$

It is very simple, using Gauss theorem and the natural symmetries of the problem, as was also noticed in [14], that the solution of this equation is:

$$\vec{j} = \frac{P_0}{4\pi r^2} \hat{r} \tag{8}$$

where $r$ is the distance from the source and where $\hat{r}$ is the unit vector $\vec{r}/r$. The RTEVI (1) is therefore not in contradiction with the inverse square low. Anyway, as this problem is a restrictive case of the second one, let us consider this other problem.

We are here working in the non-scattering / non-absorbing limit $\mu_s = \mu_a = 0$, that corresponds to standard geometrical optics, and we want to follow optical rays. We therefore look at a radiance $L(\vec{r},\vec{\Omega},t)$ peaked along a direction $\hat{u}(\vec{r})$, so that one can consider that the whole energy propagates along $\hat{u}$. In that case, we can introduce the light intensity $I(\vec{r})$, which can be defined as equal to the average diffuse intensity $\varphi(\vec{r})$. We will then have for the diffuse flux vector:

$$\vec{j} = \int \vec{\Omega} L(\vec{r},\vec{\Omega}) d\Omega \approx \int \hat{u}(\vec{r}) L(\vec{r},\vec{\Omega}) d\Omega = I(\vec{r})\hat{u}(\vec{r}) \tag{9}$$

where we have used the fact that $L(\vec{r},\vec{\Omega},t)$ is peaked along $\hat{u}(\vec{r})$, and therefore cancels if $\vec{\Omega}$ is significantly different from $\hat{u}(\vec{r})$. One can note that, while $\vec{\Omega}$ does not depend on $\vec{r}$, $\hat{u}(\vec{r})$ depends on $\vec{r}$, and there is no contradiction at this stage. If we insert equation (9) in the conservation equation, we have (omitting the source term):

$$\vec{\nabla} \cdot \vec{j} = \hat{u}(\vec{r}) \cdot \vec{\nabla} I(\vec{r}) + I(\vec{r})\vec{\nabla} \cdot \hat{u}(\vec{r}) = 0 \tag{10}$$

Before continuing, one may ask a question: does $\hat{u}(\vec{r})$ satisfies the ray optics equation? This quantity was indeed defined as a property of the radiance $L(\vec{r},\vec{\Omega},t)$, and should not correspond to geometrical optics if equation (1) is wrong … To show this point, let us multiply equation (1) by $\vec{\Omega}$, and integrate over $\vec{\Omega}$. The first term of the time-independent problem will be:

$$\int \vec{\Omega}\vec{\Omega} \cdot \vec{\nabla} L d\Omega = \vec{\nabla} \cdot \int \vec{\Omega}\vec{\Omega} L d\Omega \qquad (11)$$

where we recall that $\vec{\Omega}$ does not depend on $\vec{r}$. As $L(\vec{r},\vec{\Omega},t)$ is peaked along $\hat{u}(\vec{r})$ we can write:

$$\vec{\nabla} \cdot \int \vec{\Omega}\vec{\Omega} L d\Omega \approx \vec{\nabla} \cdot \int \hat{u}\hat{u} L d\Omega = \vec{\nabla} \cdot \left[ I(\vec{r})\hat{u}(\vec{r})\hat{u}(\vec{r}) \right]$$
$$= (\hat{u} \cdot \vec{\nabla} I)\hat{u} + I(\vec{\nabla} \cdot \hat{u})\hat{u} + I\hat{u} \cdot \vec{\nabla}\hat{u}$$

Using (10) this term therefore reduces to $I\hat{u} \cdot \vec{\nabla}\hat{u}$. From the same kind of argument, the second term reads $-2I\hat{u}\hat{u} \cdot \vec{\nabla} \ln r$. The third and last term (we have $\mu_s = \mu_a = 0$) is more complex, and we refer the reader to the appendix to show that (if $\vec{\vec{1}}$ is the identity):

$$\int \vec{\Omega}\vec{\nabla}_\Omega L d\Omega = 3\int \vec{\Omega}\vec{\Omega} L d\Omega - \vec{\vec{1}} I \approx 3\hat{u}\hat{u}I - \vec{\vec{1}} I \qquad (12)$$

Putting things together we therefore have:

$$\hat{u} \cdot \vec{\nabla}\hat{u} = \vec{\nabla} \ln n - \hat{u}\hat{u} \cdot \vec{\nabla} \ln n$$

what is exactly the ray optics equation ! We can therefore see $\hat{u}$ as the unit vector tangent to the optical ray, that is:

$$\hat{u} = \frac{\vec{\nabla}\Lambda}{n} \qquad (13)$$

Putting (13) into (10) gives (using $\hat{u} \cdot \vec{\nabla} \equiv d/ds$):

$$\frac{d}{ds} I(\vec{r}) + nI(\vec{r}) \frac{\vec{\nabla}\Lambda}{n} \cdot \vec{\nabla}(\frac{1}{n}) + \frac{1}{n} I(\vec{r})\Delta\Lambda = 0$$

or

$$\frac{1}{n}\frac{d}{ds}I(\vec{r}) + I(\vec{r})\frac{d}{ds}(\frac{1}{n}) = \frac{d}{ds}\left[\frac{I}{n}\right] = -\frac{1}{n^2}I(\vec{r})\Delta\Lambda$$

This is equation (6) we were looking for. We can note that, for a constant refractive index and a point-like source, where [14] $\Lambda(\vec{r}) = nr$, this equation reduces to:

$$\frac{dI}{ds} = -\frac{2}{r}I$$

that is plainly compatible with the inverse square law.

The RTEVI (1) therefore accounts for geometrical optics laws in the non-scattering / non-absorbing limit. Another point concerns the assertion that this equation involves the assumption of a zero divergence of the rays everywhere in the medium. To address this misunderstanding, let us reconsider the whole derivation of equation (1).

The radiance $L(\vec{r}_0, \vec{\Omega}_0, t)$ is defined so that

$$d\phi = L\, dA\, d\Omega\, \vec{\Omega}_0 \cdot \vec{u} \tag{14}$$

is the power flowing within the solid angle $d\Omega$ around $\vec{\Omega}_0$, and through the surface element $dA$ around $\vec{r}_0$ ($\vec{u}$ is a unit vector orthogonal to this surface element). The energy flow can be materialized by a vector field $\vec{\Omega}_{\vec{r}_0}(\vec{r}, \vec{\Omega}_0)$, where $\vec{\Omega}$ is a unit vector parallel at each point $\vec{r}$ to the geometrical optical ray passing through this point. Please note here that $\vec{\Omega}$ depends on $\vec{r}$, what is not the case of $\vec{\Omega}_0$. This vector field therefore represent, among a lot of rays, the part of the energy that propagates in the direction $\vec{\Omega}_0$. But how can we exactly define it, as the parallelism is not obvious with curved rays ?

The situation is summarized on figure 1 and 2: There are a lot of rays that cross the surface element $dA$, and one has to select which will be considered as parallel to $\vec{\Omega}_0$. The most immediate choice, we made in [3], is depicted in figure 1: one can simply choose the rays that are parallel to $\vec{\Omega}_0$ when crossing $dA$. But in fact, the only obligation is to have a ray parallel to $\vec{\Omega}_0$ at $\vec{r}_0$, and one can consider a linear deviation to this law when moving within $dA$, as in figure 2. One can indeed assert that such a deviation won't imply any modification in the flux $d\phi$ crossing $dA$ at the first order in $dA$, so that it won't modify the definition (14) of $L$. One should therefore be able to make this choice without any change, but now the divergence of $\vec{\Omega}$ does not cancel any more !

Let us thus reconsider the whole demonstration with this second hypothesis: the field $\vec{\Omega}$ evolves in an arbitrary way within $dA$, that is $\vec{\Omega}(\vec{r} + \delta\vec{r}) = \vec{\Omega}_0 + A\delta\vec{r}$, where the only request on the matrix $A$ is $A\vec{\Omega}_0 = \vec{0}$ (this arbitrary deviation works only for $\delta\vec{r} \perp \vec{\Omega}_0$). In the direction $\vec{\Omega}_0$, the evolution of $\vec{\Omega}$ is indeed governed by the ray optics equation.

In the following, we will introduce vector coordinates with Einstein summation convention (with $a^i = a_i$ in a Euclidean space). From former considerations we can write:

$$\partial_j \Omega^i = A_j^{\ i} + \frac{\Omega_{0j}}{n}\left[\delta^{ik} - \Omega_0^i \Omega_0^k\right]\partial_k n \qquad (15)$$

where $\delta^{ij}$ is the identity. The main difficulty, in order to obtain equation (1), is to evaluate the quantity

$$L(\vec{r}\,',\vec{\Omega}\,',t+dt)\,d\Omega'\,dA'-L(\vec{r}_0,\vec{\Omega}_0,t)\,d\Omega\,dA$$

where the primes stand for quantities at $\vec{r}\,' = \vec{r}_0 + \vec{\Omega}_0 ds$ (with $ds = c_0 dt / n$). We have now [1,3]:

$$dA' - dA = \partial_i \Omega^i \, ds \, dA = A_i^{\ i} \, ds \, dA$$

where the divergence $A_i^{\ i}$ of $\vec{\Omega}$ is a completely arbitrary number. Let us now focus on $d\Omega$; as in [3], we introduce 2 vectors $d\vec{y}_0^{(1)}$ and $d\vec{y}_0^{(2)}$ orthogonal to $\vec{\Omega}_0$ and define the solid angle $d\Omega$ as:

$$d\Omega = \vec{\Omega}_0 \cdot (d\vec{y}_0^{(1)} \times d\vec{y}_0^{(2)}) = \det(\vec{\Omega}_0, d\vec{y}_0^{(1)}, d\vec{y}_0^{(2)}) \tag{16}$$

If $\vec{\Omega}_0^{(N)} = \vec{\Omega}_0 + d\vec{y}_0^{(N)}$, we can define the vector field $\vec{\Omega}^{(N)} = \vec{\Omega}_{\vec{r}_0}(\vec{r}, \vec{\Omega}_0^{(N)}) = \vec{\Omega} + d\vec{y}^{(N)}$. Let us write the variation of $\vec{\Omega}^{(N)}$ for an infinitesimal displacement along $\vec{\Omega}_0^{(N)}$, that is:

$$\frac{d\vec{\Omega}^{(N)}}{ds} = \vec{\Omega}_0^{(N)} \cdot \vec{\nabla} \vec{\Omega}_{\vec{r}_0}(\vec{r}_0, \vec{\Omega}_0^{(N)})$$

$d\vec{\Omega}^{(N)}/ds$ is a function of $\vec{\Omega}_0^{(N)}$ that can be developed around $\vec{\Omega}_0$:

$$\frac{d\vec{\Omega}^{(N)}}{ds} = \frac{d\vec{\Omega}}{ds} + dy_0^{(N)k} \partial_{\Omega_{0k}} \frac{d\vec{\Omega}}{ds} \tag{17}$$

We can furthermore write, as $\vec{\Omega}^{(N)} \equiv \vec{\Omega} + d\vec{y}^{(N)}$:

$$\frac{d\vec{\Omega}^{(N)}}{ds} = \vec{\Omega}_0^{(N)} \cdot \vec{\nabla} \vec{\Omega} + \frac{d}{ds} d\vec{y}^{(N)} \tag{18}$$

As $d\vec{y}_0^{(N)}$ is orthogonal to $\vec{\Omega}_0$, we immediately have from (15) that

$$d\vec{y}_0^{(N)} \cdot \vec{\nabla} \Omega^i = dy_0^{(N)j} A_j^{\ i}$$

so that $\vec{\Omega}_0^{(N)} \cdot \vec{\nabla} \vec{\Omega} = \vec{\Omega}_0 \cdot \vec{\nabla} \vec{\Omega} + {}^t A \, d\vec{y}_0^{(N)} = \frac{d\vec{\Omega}}{ds} + {}^t A \, d\vec{y}_0^{(N)}$, and we can therefore conclude, by comparing (17) and (18), that:

$$\frac{d}{ds} dy^{(N)i} = dy_0^{(N)k} \partial_{\Omega_{0k}} \frac{d\Omega^i}{ds} - dy_0^{(N)j} A_j^{\ i} \tag{19}$$

and

$$dΩ' = \det \begin{bmatrix} (Ω_0^i, dy_0^{(1)i}, dy_0^{(2)i}) \\ + ds\,(\dfrac{dΩ^i}{ds}, ∂_{Ω_{0k}} \dfrac{dΩ^i}{ds} dy_0^{(1)k}, ∂_{Ω_{0k}} \dfrac{dΩ^i}{ds} dy_0^{(2)k}) \\ - ds(0, dy_0^{(1)j} A_j{}^i, dy_0^{(2)j} A_j{}^i) \end{bmatrix} \quad (20)$$

We will first notice that $d\vec{Ω}/ds$, which is orthogonal to $\vec{Ω}_0$, is in the plane $(d\vec{y}_0^{(1)}, d\vec{y}_0^{(2)})$ and do not contribute in (20) up to the first order in ds; a second remark is that, if $P^{ij} = δ^{ij} - Ω_0^i Ω_0^j$ is the projector on the plane orthogonal to $\vec{Ω}_0$, we have:

$$(0, dy_0^{(1)i}, dy_0^{(2)i}) = P_j^i (Ω_0^j, dy_0^{(1)j}, dy_0^{(2)j}) \quad (21)$$

If we recall that $A\vec{Ω}_0 = \vec{0}$, we can write the same thing with the operator $A^i{}_j$. These considerations allow us to write:

$$dΩ' = \det\left[ δ_l^i + ds\, ∂_{Ω_{0k}} \dfrac{dΩ^i}{ds} P_l^k - ds A_l^i \right] dΩ$$
$$= (1 + ds\, ∂_{Ω_{0k}} \dfrac{dΩ^i}{ds} P_i^k - ds A_i^i) dΩ \quad (22)$$

We finally obtain, after some algebra and using $P_i^i = 2$:

$$dΩ' = (1 - ds\, \dfrac{2}{n} \vec{Ω}_0 \cdot \vec{\nabla} n - ds A_i{}^i) dΩ \quad (23)$$

We therefore have, up to the first order:

$$dA'\,dΩ' = (1 + ds\, ∂_i Ω^i - ds\, \dfrac{2}{n} \vec{Ω}_0 \cdot \vec{\nabla} n - ds\, ∂_i Ω^i) dA\,dΩ$$
$$= (1 - ds\, \dfrac{2}{n} \vec{Ω}_0 \cdot \vec{\nabla} n) dA\,dΩ \quad (24)$$

where we can see that the terms containing the divergence of $\vec{Ω}$ exactly cancels, and therefore do not enter in the RTEVI (1). This divergence is arbitrary, and it was absolutely legitimate to cancel it by making the choice of figure 1.

This paper was devoted to the clarification of some points concerning the radiative transfer equation in a medium with a spatially varying refractive index (RTEVI). We have shown that the RTEVI (1) correctly describes situations in the geometrical optics limit (no absorption/ no scattering), and accounts for situations with a non-zero divergence of the optical rays.

**Appendix**

This appendix is devoted to the calculus of some integrals involving the transverse gradient $\partial_{\Omega_i} L$. Let us consider here a d-dimensional space, where $x^i$ is a vector of norm $x$ along the unit vector $\Omega^i = x^i / x$. The gradient over $\vec{x}$ of a function $f$ reads:

$$\partial_{x_i} f = \frac{\partial f}{\partial x} \Omega_i + \frac{1}{x} \partial_{\Omega_i} f \tag{A1}$$

what defines the transverse gradient in a d-dimensional space. Let us now consider the extension of the radiance $L$ as a function of $x^i$.

$$L(x^i) = L(\Omega^i = x^i / x) \tag{A2}$$

As $L$ does not depend on the norm $x$, one has from (A1):

$$\frac{\partial L}{\partial x^i} = \frac{1}{x} \frac{\partial L}{\partial \Omega^i} \tag{A3}$$

Let us now integrate on the spherical shell of figure 3:

$$\int \frac{\partial L}{\partial x^i} d^d x = \int \frac{\partial}{\partial x^j} \left[ \delta^j{}_i L \right] d^d x \tag{A4}$$

where $d^d x = x^{d-1} d\Omega dx$. The first term in equation (A4) is:

$$\int \frac{\partial L}{\partial x^i} d^d x = \int x^{d-2} dx \int \frac{\partial L}{\partial \Omega^i} d\Omega$$

$$= \frac{1}{d-1} \left[ X_2^{d-1} - X_1^{d-1} \right] \int \frac{\partial L}{\partial \Omega^i} d\Omega$$

The second term, in the right of equation (A4), can be evaluated using Gauss theorem, where the outward unit normal field of the bigger boundary sphere is $\Omega^i$, while it is $-\Omega^i$ for the smallest one (see figure 3):

$$\int \frac{\partial}{\partial x^j}\left[\delta^j{}_i L\right]d^d x = \int_{S(X_2)}\left[\delta^j{}_i L\right]\Omega_j d^{d-1}x - \int_{S(X_1)}\left[\delta^j{}_i L\right]\Omega_j d^{d-1}x$$

$$= [X_2^{d-1} - X_1^{d-1}]\int_{4\pi}\Omega_i L(\vec{\Omega})d\Omega$$

Equation (A4) therefore reads:

$$\int \frac{\partial L}{\partial \Omega^i}d\Omega = (d-1)\int_{4\pi}\Omega_i L(\vec{\Omega})d\Omega$$

In dimension $d=2$, this corresponds to equation (5). The same work can be performed on:

$$\int x^j \frac{\partial L}{\partial x^i}d^d x = \int \frac{\partial}{\partial x^i}\left[x^j L\right]d^d x - \int \delta^j{}_i L d^d x$$

$$= \int \frac{\partial}{\partial x^k}\left[\delta^k{}_i x^j L\right]d^d x - \int \delta^j{}_i L d^d x$$
(A5)

We apply the same procedure: the first term is

$$\int x^j \frac{\partial L}{\partial x^i}d^d x = \frac{1}{d}\left[X_2^d - X_1^d\right]\int \Omega^j \frac{\partial L}{\partial \Omega^i}d\Omega$$

while

$$\int \frac{\partial}{\partial x^k}\left[\delta^k{}_i x^j L\right]d^d x = \left[X_2^d - X_1^d\right]\int \Omega_i \Omega^j L d\Omega$$

and

$$\int \delta^j{}_i L d^d x = \frac{1}{d}\left[X_2^d - X_1^d\right]\delta^j{}_i \int L d\Omega$$

so that we have:

$$\int \Omega^j \frac{\partial L}{\partial \Omega^i}d\Omega = d\int \Omega_i \Omega^j L d\Omega - \delta^j{}_i \int L d\Omega$$

what gives exactly equation (12) in dimension $d=3$.

**Figure caption**

Figure 1: Rays passing through the surface element $dA$. In bold are the rays implied in the definition of $L(\vec{r}_0,\vec{\Omega}_0,t)$: in this figure, these rays are parallel to $\vec{\Omega}_0$ when crossing $dA$.

Figure 2: Same as figure 1, but now the bold rays (chosen for the definition of $L(\vec{r}_0,\vec{\Omega}_0,t)$ ) present a linear deviation versus the displacement $\vec{\delta r}$ within $dA$.

Figure 3: The integration volume considered is a spherical shell, with boundaries $x = X_1$ and $x = X_2$.

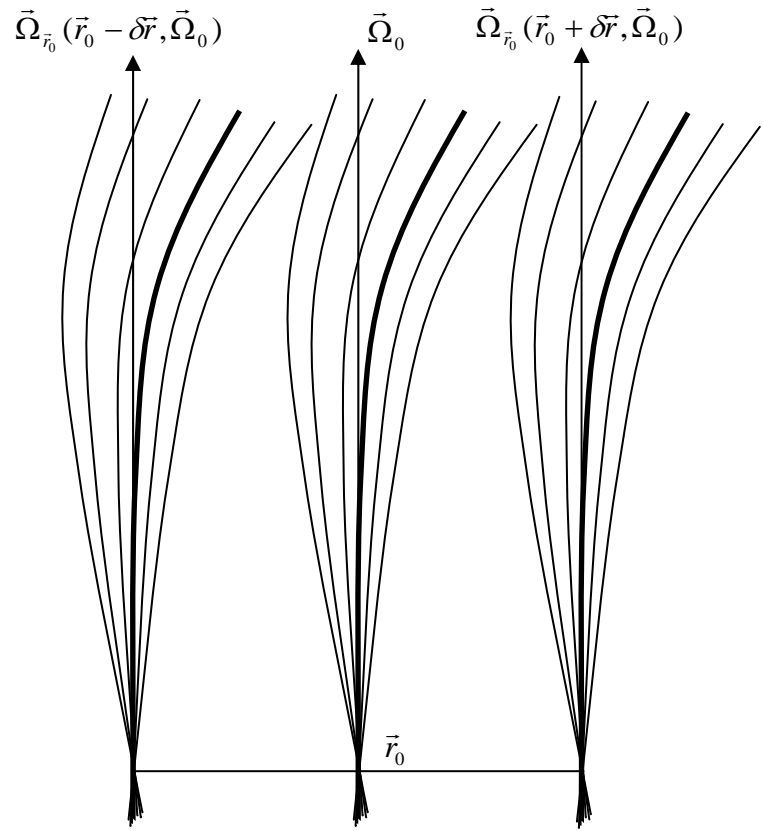

Figure 1

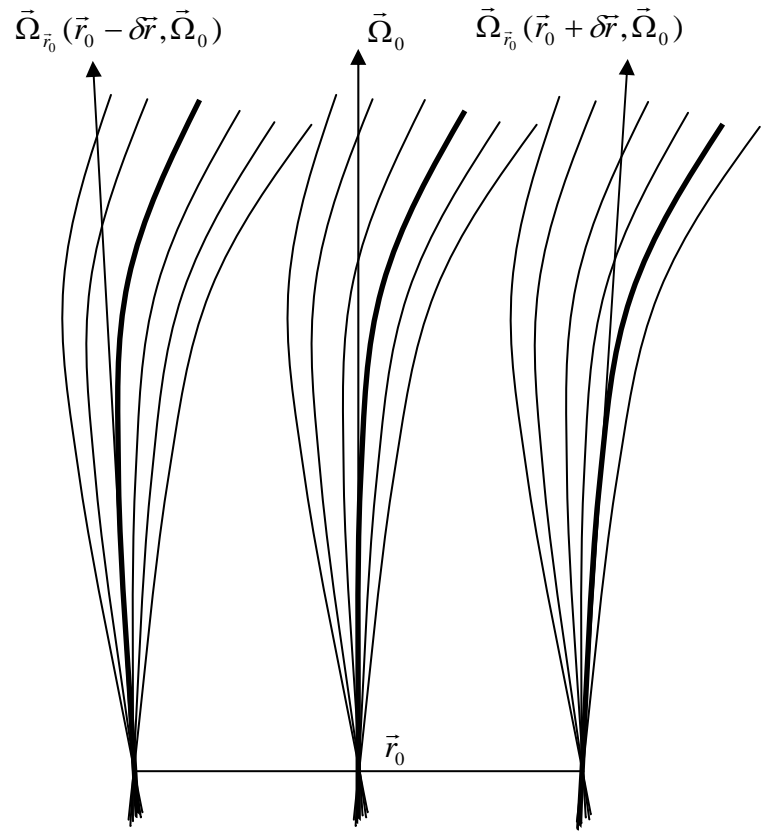

Figure 2

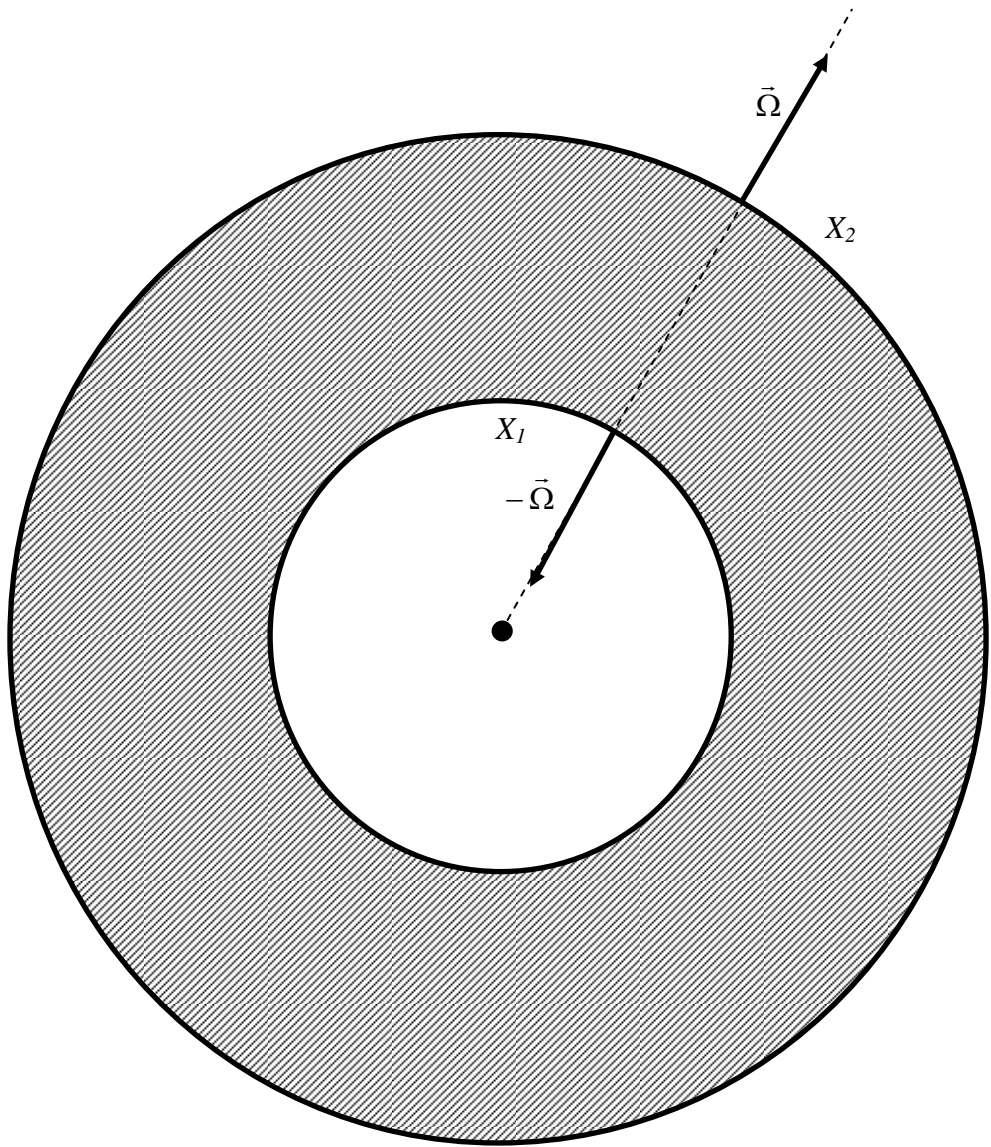

Figure 3